# LATTICE QCD DATA AND METADATA ARCHIVES AT FERMILAB AND THE INTERNATIONAL LATTICE DATA GRID*

E. Neilsen,[#] J. Simone, FNAL, Batavia, IL 60510, USA

*Abstract*

The lattice gauge theory community produces large volumes of data. Because the data produced by completed computations form the basis for future work, the maintenance of archives of existing data and metadata describing the provenance, generation parameters, and derived characteristics of that data is essential not only as a reference, but also as a basis for future work. Development of these archives according to uniform standards both in the data and metadata formats provided and in the software interfaces to the component services could greatly simplify collaborations between institutions and enable the dissemination of meaningful results.

This paper describes the progress made in the development of a set of such archives at the Fermilab lattice QCD facility. We are coordinating the development of the interfaces to these facilities and the formats of the data and metadata they provide with the efforts of the international lattice data grid (ILDG) metadata and middleware working groups, whose goals are to develop standard formats for lattice QCD data and metadata and a uniform interface to archive facilities that store them. Services under development include those commonly associate with data grids: a service registry, a metadata database, a replica catalog, and an interface to a mass storage system. All services provide GSI authenticated web service interfaces following modern standards, including WSDL and SOAP, and accept and provide data and metadata following recent XML based formats proposed by the ILDG metadata working group.

## INTRODUCTION

Lattice quantum chromodynamics (QCD) is the numerical simulation of the strong interaction between quarks using discretized space and time. Lattice QCD projects generate sets of large files representing gluon states through a sequence of simulation times. Several groups generate and archive such data. These data are vital resources both for other groups and for future use by the original authors.

The discovery, distribution, and use of this archived data presents several challenges. The data volume is inconveniently large, so the data is commonly stored in mass storage systems, and must be transferred between mass storage systems at different sites to be used conveniently at each site. The metadata available electronically is often incomplete, and that which exists is usually not conveniently searchable; discovering data of interest and interpreting it properly often requires contacting the scientists who generated the data personally.

## THE ILDG

The International Lattice Data Grid (ILDG)[1] will provide infrastructure to help meet these challenges.

When complete, the ILDG will consist of an integrated set of clients and services, implemented (possibly independently) at each participating institution, that enables users to make data available to other scientists and to discover and use data made available by others.

This infrastructure should have at least the following functionality:

*Export*

A user must be able to make data available.

*Explore*

A user should be able to interactively explore available data using an interface that presents an intuitive, hierarchical sorting of the data based on metadata.

*Discover*

A user must be able to list data whose metadata matches desired criteria.

*Query*

A user should be able to retrieve metadata about a specific set of data, either specific parameters or full files in a standard format.

*Replicate*

A user must be able to copy a data file from one mass storage system into another when access to both is permitted.

*Get*

A user must be able to copy a data file and corresponding metadata onto local disk.

___

*Alter*

A user should be able to publish revisions to the metadata. (The data itself should not be alterable.)

*Revert*

It should be possible to revert to a previous version of the metadata in case of accidental corruption.

*Audit*

It should be possible to track usage based on a variety of parameters.

*Withdraw*

A user should be able to withdraw data from the ILDG. This operation will not make either the data or metadata unavailable, but it should prevent the file from being listed by metadata queries unless withdrawn data is specifically requested.

## POINTS OF COLLABORATION

In order to share data effectively, participants must be able both to use data grid services provided by other groups and to read and interpret the data correctly once it is obtained from those services. These requirements do not necessitate the use the same software by all institutions; common interfaces to the services, sufficiently complete metadata, and common formats for data and metadata are sufficient.

The ILDG collaboration recently specified version 1.1 of QCDml[2], an XML based metadata format for lattice QCD, and is working on a standard for storing the lattice data itself.

The ILDG middleware working group is generating standard middleware interfaces for the desired grid services following globus and World Wide Web Consortium (W3C) recommended standards. The use of these standards will allow developers implementing data grid clients and services to take advantage of existing code libraries, and even to enable ambitious users to develop software that contacts the services directly. The user interfaces for the clients need not be specified to ensure interoperability, and may be customized to simplify local usage patterns.

## DATA GRID SERVICES

The ILDG follows an architecture typical of data grids: users (or their client software) discover available services using a web-service registry, discover data and obtain necessary metadata using a metadata database, locate data using a replica catalog, and manage data movement using a storage resource manager.

ILDG clients interact with the various services through the exchange of Simple Object Access Protocol (SOAP) messages. Most interactions will begin with a client sending a web-service register SOAP message requesting a list of a particular type of service, and the registry service returning a SOAP message with an appropriate list. Similarly, a client wishing to query a metadata database will send a SOAP message to the desired metadata database web-service with the desired query, and receive a SOAP message from the web-service with the response to the query.

The services perform most of the work of querying the database itself, moving the data, etc.; the client is only there to orchestrate the activities of the various services to carry out the operations specified by the user.

## THE FERMILAB PROTOTYPE

The ILDG middleware working group has not yet completed a set of interfaces, but is working to finalize the details of the requirements and architecture. However, the participating institutions are generating prototypes of the services according to the current draft documents in order to uncover deficiencies in the design and experiment with different interface options.

We, the ILDG middleware working group participants at Fermilab, have generated prototypes for a client and each of the services.

We use the Fermilab implementation of SRM, a service implemented outside of the ILDG project and in production use in other projects, as our storage resource manager. We implemented the client and remaining services in Python, using globus and pyGlobus to support grid security infrastructure (GSI) authenticated remote procedure calls using SOAP. We store replica information and metadata in a postgreSQL database managed by our python applications.

The Fermilab prototype supports most of the requisite functionality; although it implements each use in a simple, no frills manner, only the "audit" functionality is inadequate for real work. The client integrates the use of the services, providing a simple interface to the user and removing the need for a user to handle the details of the interface. For example, although the services consistently refer to data sets using global file names, the user need never be aware that global file names even exist.

Although the services themselves are running and implement enough functionality to be useful, the databases currently contain little data; until metadata and replica information for a significant quantity of existing data can be generated and loaded into the database, the usefulness of the prototype will be severely limited.

## THE METADATA CATALOG

The metadata database service serves as an interface to a database holding the metadata describing the data distributed by the ILDG. Simple, intuitive querying and browsing of this database is essential.

Because data is moved into and out of the metadata database in XML format, is it tempting to use an XML

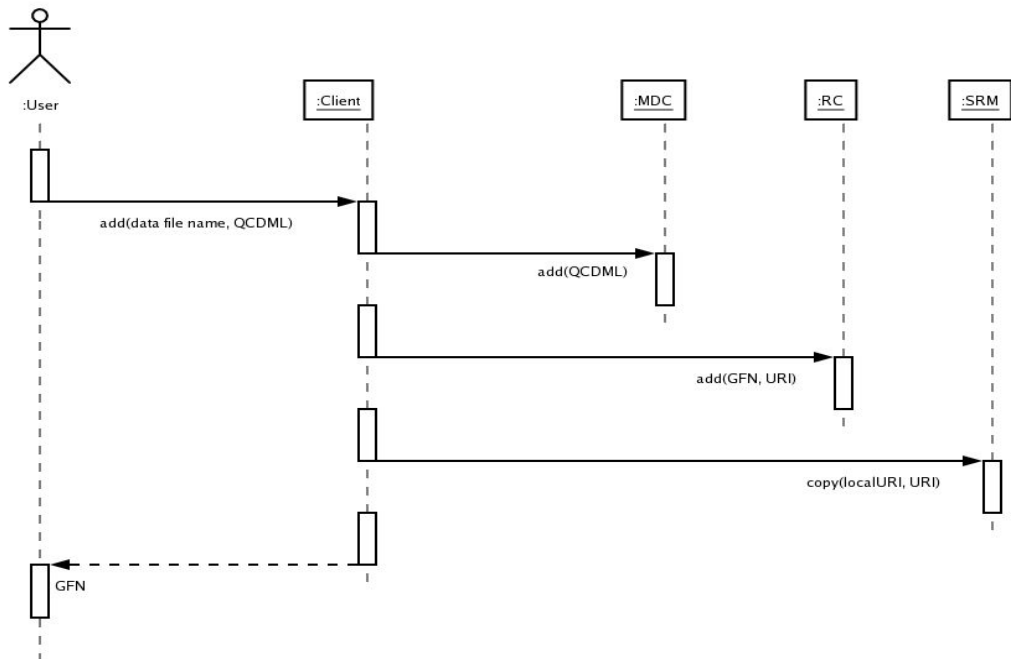

Figure 1: Interaction Diagram for the Export Use Case

database to store it. However, the native XML database query languages we explored were awkward to use interactively, and the databases themselves appear not to scale well.[3]

While naive automatic mappings of XML onto a relational database are possible, they are not optimal; we preferred a hand designed, normalized database for our prototype. The prototype allows the user to query a single SQL view of the database, hiding the complexity of the database schema. In most but not all cases, an XML path in one of the QCDml files corresponds directly to a column in this view.

## SAMPLE INTERACTIONS

### Export Ensemble

The ILDG metadata file format defines two flavors of metadata file: ensemble metadata, which applies to a set of many data files, and configuration metadata, which applies to a single data file. The complete metadata for any individual file includes both of these files.

Before new data may be submitted to the Fermilab ILDG prototype, the ensemble of which the data file is a member must be defined. The metadata file defining the ensemble can be entered into the metadata database using the following command:
$ ildg export-ensemble mydata1.qcdml

Figure 2 shows the interactions between the client and each of the services during the execution of this command.

### Export Data

Once ensemble metadata has been entered into the metadata database, individual files and corresponding metadata may be added. In the Fermilab ILDG prototype, this may be done using the following command:
$ ildg add /data/myData.dat /data/myData.xml \
> srm://myrobothost.mylab.org/myDirectory/myData.dat

Figure 1 displays the interactions between the client and each of the services during the execution of this command.

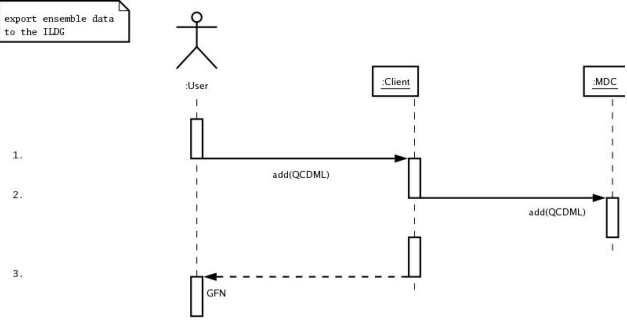

Figure 2: Interaction Diagram for the Export Ensemble Use Case

### Discover

A user can search for data using the metadata database using either a command line query or through a web browser interface. An example of a command line query is shown here:
$ ildg discover \
> "institution = 'Fermilab' AND date = '2003-12-03'"

The resulting interactions are shown in figure 3.

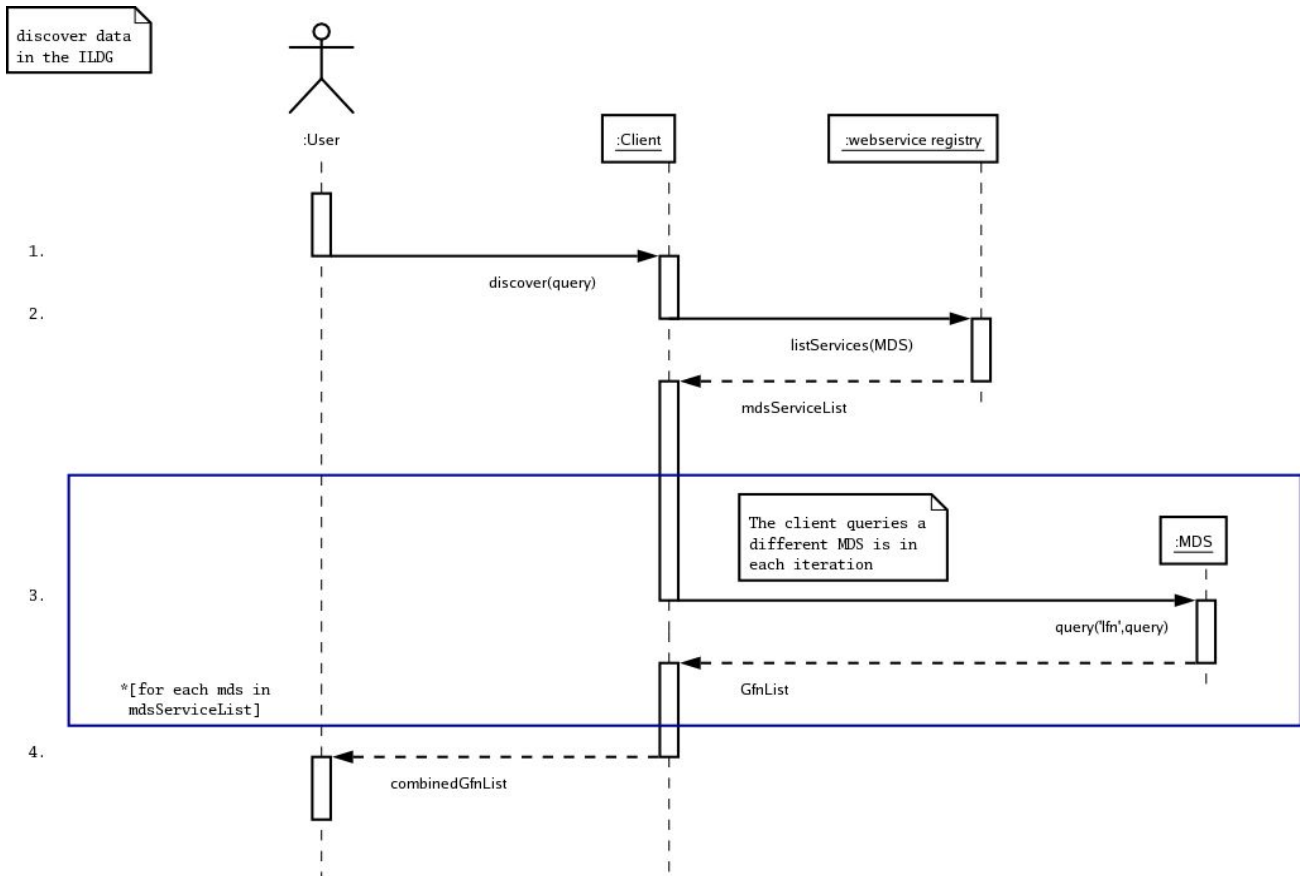

**Figure 3: Interaction Diagram for the Discover Use Case**

*Obtaining Data and Metadata*

Once the user has designed a query that selects the files desired, the following commands create local copies of the desired files:
$ ildg get "projectName = 'myProject'" \
> '/data/%s-series%s-step%s.dat' \
> 'projectName series updates'
$ ildg query "projectName = 'myProject'" \
> '/data/%s-series%s-step%s.xml' \
> 'projectName series updates'

Note that the names of the local files are generated automatically from the metadata according to the users specifications; the user need never encounter the file names used by any other entity (such as the replica catalog or archival system).

## FUTURE DIRECTIONS

Before any implementation of the ILDG software to be useful, metadata files for the existing data must be generated. Because much of the information to be included in these files is not currently in electronic form, the is a significant task requiring human attention. Once these file exists, they will be loaded into the prototype, which can then be tested by select users in a more realistic environment.

For the full vision of the lattice data grid to be realized, however, several additional tasks are required. Collaborators must reach consensus on the full details of the functionality, define standard interfaces for each service, and generate production implementations of these services and any clients local users desire.

## ACKNOWLEDGEMENTS

The work presented in this paper is based on discussions and documents generated by the ILDG middleware working group, convened by Balint Joo, Mitsuhisa Sato, and Chip Watson.